\documentclass[11pt]{article}
\usepackage{moriond,epsfig}

\def\be{\begin{equation}}
\def\ee{\end{equation}}
\def\bea{\begin{eqnarray}}
\def\eea{\end{eqnarray}}

\def\beq{\begin{equation}}
\def\eeq{\end{equation}}
\def\C{{\cal C}}

\def\d{{\delta}}
\def\E{{\cal E}}
\def\H{{\cal E}}
\def\A{{\cal A}}

\makeatletter
\@ifundefined{lesssim}{\def\lesssim{\mathrel{\mathpalette\vereq<}}}{}
\@ifundefined{gtrsim}{\def\gtrsim{\mathrel{\mathpalette\vereq>}}}{}
\def\vereq#1#2{\lower3pt\vbox{\baselineskip1.5pt \lineskip1.5pt
\ialign{$\m@th#1\hfill##\hfil$\crcr#2\crcr\sim\crcr}}}
\makeatother

\begin{document}
\vspace*{4cm}
\title{COSMOLOGY WITH AN EXTRA-DIMENSION}

\author{David Langlois}

\address{GReCO, Institut d'Astrophysique de Paris, \\
98bis Boulevard Arago, 75014 Paris, France}

\maketitle\abstracts{
I present  the main features of the brane cosmology scenario, 
according to  which our universe is a self-gravitating brane 
 embedded in 
a five-dimensional spacetime. }

\section{Introduction}
The purpose of this contribution is to give an introduction to a new 
vision of extra-dimensions: the so-called ``braneworld'' picture.
In this scenario, ordinary matter is trapped in a three-dimensional 
space, called `brane',  embedded in a higher dimensional space.
 
This  idea must be contrasted with the traditional view of extra dimensions,
the Kaluza-Klein picture, where matter fields live {\it everywhere}
 in {\it compact} extra dimensions.  Any such 
higher dimensional field can  be described, via a  Fourier expansion, 
 as an infinite collection of 
 four-dimensional fields, the so-called Kaluza-Klein modes, with  
masses depending on the size of the extra-dimensions. Non-observation 
of Kaluza-Klein modes in colliders thus gives an upper bound on the size 
of the extra-dimension, 
typically 
\beq
R\lesssim E_{max}^{-1}, 
\eeq
 where $E_{max}$ is the highest energy probed in colliders, i.e.  
$E_{max}\sim 1$ TeV today.

In the context  of string theory, extra dimensions were promoted from 
an interesting curiosity to a theoretical necessity since  superstring theory 
requires ten space-time dimensions  to be consistent from the quantum 
point of view. Until recently, the 
six extra-dimensions were however  hidden, or 
`compactified', {\it \`a la} Kaluza-Klein. The new implementation of 
extra dimensions has emerged lately, notably 
with M-theory and the Horava-Witten 
model \cite{hw} which describes the low 
energy effective theory corresponding to the strong coupling limit 
of $E_8\times E_8$ heterotic string theory. 
This model gives  an  eleven-dimensional  bulk spacetime 
with 11-dimensional supergravity, 
 the eleventh dimension being compactified via a  $Z_2$ orbifold symmetry.
The two fixed points of the orbifold symmetry define two 10-dimensional 
spacetime boundaries, or 9-branes, on which the gauge groups are defined.
A dimensional reduction of this model to five space-time dimensions yields 
a picture with two $Z_2$-symmetric three-branes separated in a fifth 
dimension, a picture somewhat similar to the framework of brane cosmology 
presented below.

\section{Homogeneous cosmology in a brane-universe}

The main motivation for exploring cosmology in models with extra-dimensions 
is that the  signature of extra-dimensions might be accessible 
{\it only} at very high energies, i.e. in the very early universe.  
One would thus like to investigate what kind 
of  relic imprints  could be left and  tested today via
 cosmological observations.
The purpose of this section is to describe the  homogeneous cosmology 
\cite{bdl99,bdel99,ftw99}
of a self-gravitating brane-universe embedded in a five-dimensional 
spacetime.

Let us thus consider a five-dimensional spacetime with three-dimensional
isotropy and homogeneity, which contains a three-brane representing
our universe. It is convenient, but not necessary, 
 to work in a Gaussian normal coordinate
system based on  our brane-universe. Due to the spacetime 
symmetries, the metric is then of the form
\beq
ds^2=- n(t,y)^2 dt^2+a(t,y)^2 \delta_{ij}dx^idx^j+dy^2,
\label{metric}
\eeq
where we have assumed that our brane-universe is spatially flat (but 
this can be generalized very easily to hyperbolic or elliptic spaces).
In these coordinates, our brane-universe is always located at $y=0$.

The energy-momentum tensor can be decomposed into a bulk 
energy-momentum tensor, which will be assumed to vanish here, 
 and a brane energy-momentum tensor, the latter 
being of the form
\beq
 T^A_{\, B}= S^A_{\, B}\delta (y)= \{\rho_b, p_b, p_b, p_b, 0\}\delta (y),
\eeq
with a delta function since ordinary  matter  is confined 
in the brane. $\rho_b$ and $P_b$ are respectively 
the total energy density and pressure in the brane and depend only
on time.
Allowing  for a cosmological 
constant in the bulk, $\Lambda$,   the 
five-dimensional Einstein equations read
\beq
G_{AB}+\Lambda g_{AB}=\kappa^2 T_{AB}.
\label{einstein}
\eeq
In order to determine the  influence of the brane 
on the curvature of space-time, it is useful to 
 integrate Einstein's equations in the vicinity of the brane, which leads to 
 the  so-called junction conditions  for the metric
 at the brane location $y=0$. 
According to these junction conditions,   the metric must be 
continuous and  the jump of the extrinsic curvature tensor 
$K_{AB}$  (related
to the derivatives of the metric with respect to $y$) depends on 
the distributional energy-momentum tensor, 
 \beq
\left[K^A_{\, B}
-K\delta ^A_{\, B}\right]=\kappa^2 S^A_{\, B},
\label{israel}
\eeq
where the brackets here denote the jump at the 
brane, i.e. $[Q]=Q_{\{y=0^+\}}-Q_{\{y=0^-\}}$, and the extrinsic curvature 
tensor is defined by 
$K_{AB}=h_{A}^C\nabla_C n_B$,
$n^A$ being the unit vector normal to the brane and $h_{AB}=g_{AB}-n_An_B$
the induced metric.
Assuming  that the brane is 
mirror symmetric implies
 that the jump in the extrinsic curvature is twice its value on one 
side. 
Substituting  the ansatz metric (\ref{metric}) in (\ref{israel}), one ends up 
with the two junction conditions \cite{bdl99}:
\beq
\left({n'\over n}\right)_{0^+}={\kappa^2\over 6}\left(3p_b+2\rho_b\right),
\qquad
\left({a'\over a}\right)_{0^+}=-{\kappa^2\over 6}\rho_b.
\label{junction}
\eeq
 
Solving explicitly\cite{bdel99} 
the Einstein equations 
(\ref{einstein}) with the metric ansatz (\ref{metric}), one finds in particular
that the geometry induced 
{\it in the brane} is governed by 
 the  equation 
\beq
H_0^2\equiv {\dot a_0^2\over a_0^2}={\kappa^4\over 36}\rho_b^2+{\Lambda\over 6}
+{\C\over a^4}.
\label{fried}
\eeq
where the subscript `$0$' means evaluation at $y=0$, i.e. $a_0(t)\equiv 
a(t,y=0)$ and $\C$ is an integration constant.
This equation is analogous to the  (first)
Friedmann equation, since it relates the Hubble parameter to the 
energy density, but  is nevertheless different [the usual Friedmann equation
 reads $H^2=(8\pi G/3)\rho$]. 
 Its most  remarkable feature  is that the energy density of the brane enters 
{\it }quadratically on the right hand side in contrast with the standard 
four-dimensional Friedmann equation where the energy density enters 
linearly. 
Another consequence of the five-dimensional Einstein equations 
(\ref{einstein}) is that 
the  energy conservation equation  
is unchanged 
and still reads
\beq
\dot\rho_b+3H(\rho_b+p_b)=0. 
\label{conserv}
\eeq

In the simplest case where $\Lambda=0$ and $\C=0$, 
one can easily solve the above cosmological equations 
(\ref{fried}-\ref{conserv}) for 
a perfect fluid with an equation of state $p_b=w\rho_b$ and $w$ constant.
One finds that the evolution of the scale factor is given by
\beq
a_0(t)\propto t^{1\over 3(1+w)}.
\label{hesf}
\eeq
In the most interesting cases for cosmology, radiation and pressureless 
matter, one finds respectively  
$a\sim t^{1/4}$  (instead 
of the usual $a\sim t^{1/2}$) and $a\sim t^{1/3}$  (instead 
of  $a\sim t^{2/3}$).
Such behaviour  is problematic because it cannot be reconciled 
with nucleosynthesis. Indeed, the nucleosynthesis scenario depends on the 
balance between   the microphysical  reaction rates
 and the expansion rate of the universe, and changing in a drastic way 
the evolution of the scale factor between nucleosynthesis and today
 modifies dramatically the predictions for the light element abundances.

The modified  Friedmann law (\ref{fried}), with the $\rho_b^2$ term
 but without the 
bulk cosmological constant (and without the $\C$ term) was  derived  
\cite{bdl99} just before  a new model describing a flat (Minkowski) world 
with one extra-dimension was proposed by Randall and Sundrum\cite{rs99b}. 
The new ingredient 
was to endow  our brane-world  with a tension (constant energy 
density) and the five-dimensional bulk with a {\it negative} cosmological 
constant, the two being fine-tuned so that the effective four-dimensional 
Hubble parameter is zero in (\ref{fried}) (taking $\C=0$). It turns out that 
such a set-up gives the usual four-dimensional gravity \cite{rs99b,gt99} 
(except on very small 
scales).  

This  suggested that  the  generalization of the Randall-Sundrum 
model  to cosmology   should  be compatible with standard cosmology 
at small energy 
scales, as it was shown \cite{cosmors} speedily.
If one wants to go beyond a Minkowski geometry  and consider 
non trivial   cosmology in the brane, one must  assume 
that the total energy density 
in the brane, $\rho_b$, consists of two parts, 
\beq
\rho_b=\sigma+\rho,
\eeq
the tension $\sigma$, constant in time, and the usual cosmological energy
density $\rho$. 
Substituting this decomposition into (\ref{fried}), one obtains 
\beq
H^2= \left({\kappa^4\over 36}\sigma^2+{\Lambda\over 6}\right)
+{\kappa^4\over 18}\sigma\rho
+{\kappa^4\over 36}\rho^2+{\C\over a^4}.
\label{friedrs}
\eeq
Let us now fine-tune the brane tension and the bulk  cosmological constant so
that the first term on the right hand side vanishes.
The second term then becomes the dominant  term if $\rho$ is small enough and
{\it one thus recovers the usual Friedmann equation at low energy}, 
with the identification
\beq
8\pi G= {\kappa^4\over 6}\sigma,
\label{newton}
\eeq
which also agrees with Newton's constant deduced from gravitational 
interaction between test masses. 
The third term on the right hand side of (\ref{friedrs}), 
quadratic in the energy density, 
provides a {\it high-energy correction} to the Friedmann equation 
which becomes significant when the value of the energy density approaches 
the value of the tension $\sigma$ and dominates  at higher 
energy densities. In the very high energy regime, $\rho\gg \sigma$, one 
thus recovers the unconventional behaviour of (\ref{hesf}) since the 
bulk cosmological constant becomes negligible.
 For an equation 
of state $p=w\rho$, with $w$ constant,  the 
conservation equation (\ref{conserv}) gives  as usual
\beq
\rho=\rho_0 a^{-q}, \quad q\equiv 3(w+1),
\eeq
which after substitution in the Friedmann equation (\ref{friedrs}) yields
(for $\C=0$)
\beq 
a(t)=\left[q m_0 t\left(1+{q\over 2}\mu t\right)\right]^{1/q}, 
\eeq 
where we have introduced the two mass scales 
\beq
m_0\equiv  {\kappa^2\over 6}\rho_0, \qquad \mu\equiv \sqrt{-\Lambda/6}.
\eeq
 One sees that the solution for the scale 
factor interpolates between the low energy regime and the 
high energy regime and that the transition time is of the order of $\mu^{-1}$, 
which is the characteristic mass scale associated with the cosmological 
constant.

Finally, the last term on  the right hand side of (\ref{friedrs}) 
behaves like radiation and 
arises from the integration constant $\C$. This 
constant $\C$ is quite analogous to the Schwarzschild mass and it is  
related to the bulk Weyl tensor, which vanishes when $\C=0$. In a 
cosmological context, this term is constrained to be small enough 
at the time of nucleosynthesis in order to satisfy the constraints on the 
number of extra light degrees of freedom. 

 The metric  outside the brane can be also determined explicitly
\cite{bdel99}.
In the special  case $\C=0$, the metric has a very simple form and 
its components are given by   
\begin{eqnarray}
a(t,y)&=& a_0(t)\left(\cosh\mu y-\eta \sinh\mu|y|\right)\\
n(t,y)&=& \cosh\mu y-\tilde\eta \sinh\mu|y|
\label{bulk_metric}
\end{eqnarray}
where
\beq
\eta=1+{\rho\over\sigma}, \qquad \tilde\eta=\eta+{\dot\eta\over H_0}
\eeq
and we have chosen the time $t$ corresponding to the cosmic time in the 
brane. 
The Randall-Sundrum model  corresponds to 
 $\rho=0$, i.e. $\rho_b=\sigma$, which 
implies  $\eta=\tilde\eta=1$ and one recovers
$a(t,y)=a_0\exp(-\mu|y|)$.

As explained above, the Randall-Sundrum  version of brane cosmology
gives at sufficiently late times a cosmological evolution identical 
to usual four-dimensional cosmology. One must simply ensure that 
the low-energy regime  encompasses the periods that are well constrained
by cosmological observations. 
This essentially means  that 
nucleosynthesis must take place in the low-energy regime. 
This is the case if 
the energy scale associated with the tension is higher than 
the nucleosynthesis
energy scale, i.e.
\beq
\sigma^{1/4} \gtrsim 1 \ {\rm MeV}.
\eeq
Combining this with (\ref{newton}) implies for the fundamental mass scale
(defined by $\kappa^2=M^{-3}$)  
\beq
M \gtrsim 10^4 \ {\rm GeV}. \qquad [{\rm nucleosynthesis}]
\eeq
There is however another constraint, which is not of cosmological nature:
the requirement to recover ordinary gravity down to scales of the 
submillimeter order, which have been probed by gravity 
experiments\cite{grav_exp}. This implies
\beq
\ell=\mu^{-1} \lesssim  10^{-1} \ {\rm mm},
\eeq
which yields the constraint
\beq
 M \gtrsim 10^8 \ {\rm GeV} \qquad[{\rm gravity \,  experiments}].
\eeq
The most stringent constraint is thus the latter, independent from cosmology.
Even in the low-energy regime, there is an additional constraint
 restricting the range of  values for the Weyl parameter $\C$. There are 
indeed bounds on the number of additional 
relativistic degrees of freedom allowed during nucleosynthesis (usually 
expressed as the number of additional light neutrino species). 
Typically, this gives the constraint 
\beq
{\rho_{Weyl}\over \rho_{rad}}\equiv {\C\sigma\over 2 a^4\mu^2\rho} \lesssim
10 \%.
\eeq

So far, we have thus been able to build a model, which reproduces all 
qualitative and quantitative features of ordinary cosmology in the 
domains that have been tested by observations. 
The obvious next question is whether this will still hold for a more 
realistic cosmology that includes  perturbations  from homogeneity, and more 
interestingly, whether  
brane cosmology is capable of 
providing  predictions that  deviate from usual cosmology and 
which might tested in the  future. 
A few elements are given in the next section but the main answer is still 
unkown.

\section{Brane cosmological perturbations}
Endowed with a viable homogeneous scenario, 
one would like to explore the much richer domain 
of cosmological perturbations and investigate whether brane cosmology 
leads to new effects that could be tested in the forthcoming 
cosmological observations, in particular of 
the anisotropies of the Cosmic Microwave Background.

Brane cosmological perturbations is a difficult subject and although there 
are now many published works  on this question, 
no observational  signature has yet been  
 predicted. 
Below I will summarize some results concerning two differents aspects 
of perturbations. The first aspect deals with the evolution of 
scalar type perturbations on the brane, the second aspect with the 
production of gravitational waves from quantum fluctuations during a 
de Sitter phase in the brane.

Let us first  discuss scalar type cosmological perturbations in brane 
cosmology. 
Choosing  Gaussian normal 
 coordinates, the metric  with scalar perturbations can 
be written as \cite{l00a,l00b}
\beq
ds^2=-n^2 (1+2 A)dt^2+2n^2 \partial_i B dt dx^i 
+a^2\left[ (1+2C)\d_{ij}+2\partial_i \partial_j E\right]dx^i dx^j
+dy^2,
\eeq
where the perturbations at $y=0$ turn out to coincide exactly with 
 the standard scalar cosmological perturbations. 
 Using the compact notation 
$h_\alpha=\{A,B,C,E\}$ ($\alpha=1,\dots, 4$),
the linearized Einstein equations 
\beq
\delta G_{AB}+\Lambda \d g_{AB}=\kappa^2 \d T_{AB}
\eeq
yield, {\it in the bulk}, expressions of the form
\begin{eqnarray}
\delta G^{(5)}_{00}&=&  \delta 
G^{st}_{00}+
[h_\alpha, h'_\alpha,h''_\alpha] =-\Lambda\delta g_{00}\\
\delta G^{(5)}_{ij} &=& \delta G^{st}_{ij}+
[h_\alpha, h'_\alpha,h''_\alpha]=-\Lambda\delta g_{ij}\\
\delta G^{(5)}_{0i} &=&  \delta G^{st}_{0i}+
[h_\alpha, h'_\alpha,h''_\alpha] =-\Lambda\delta g_{0i} \\
\delta G^{(5)}_{05} &=&  [h_\alpha, h'_\alpha]=0  \\
\delta G^{(5)}_{5i} &=&  [h_\alpha, h'_\alpha] =0\\
\delta G^{(5)}_{55} &=&  [h'_\alpha, h''_\alpha]=0
\end{eqnarray}
where the  brackets stand for linear combinations of the perturbations 
and their derivatives.
Brane matter enters only in the junction conditions, which at the linear 
level relate the first derivatives (with respect to $y$) of the 
metric perturbations $h'_\alpha$ to brane matter perturbations 
$\delta\rho$, $\delta P$, $v$, $\pi$. One can then substitute these 
relations back into the perturbed Einstein equations. 
$\delta G^{(5)}_{05}=0$ then yields the usual perturbed  
energy  conservation equation, whereas 
$\delta G^{(5)}_{i5}=0$ yields the perturbed Euler equation.
The other equations yield equations of motion for the perturbations 
where one recognizes the usual equations of motion in ordinary cosmology, 
but with  two  types of corrections \cite{l00b}:
\begin{itemize}
\item modification of the  homogeneous background coefficients due to the 
additional terms in the Friedmann equation.   
These  corrections are  negligible in the low
energy  regime 
$\rho\ll\sigma$.
For long wavelength (larger than the Hubble scale) perturbations, 
one can thus obtain  a transfer coefficient, $T=5/6$, characterizing the 
high/low energy   transition, i.e. 
\beq
\Phi_{\rho\ll\sigma}={5\over 6}\Phi_{\rho\gg\sigma}.
\eeq
\item presence of source terms in the  equations. 
These terms come from the bulk perturbations and cannot be determined solely 
from the evolution inside the brane. To determine them, one must solve 
the full problem in the bulk (which also means to specify some initial 
conditions in the bulk). From the four-dimensial point of view, these 
terms from the fifth dimension appear like external source terms and their 
impact is formally similar to that of ``active seeds'', which have been 
studied in the context of topological defects. 
\end{itemize}

Let us  now discuss the origin of  brane cosmological perturbations.
 In standard cosmology, the main mechanism for producing 
the cosmological perturbations is inflation. This can be
generalized  to  brane cosmology, for instance via  
 a scalar field confined in the brane\cite{mwbh99}. 
While the scalar perturbations depend on the physical details of the 
inflationary model, gravitational waves are essentially sensitive
only to the geometry and, in this sense, their predictions are more robust.
Let us thus consider here  the spectrum of gravitational waves generated
during brane inflation\cite{lmw00}.   

To compute the production of gravitational waves, one 
can approximate slow-roll brane inflation by a succession of de Sitter phases.
The metric for a de Sitter brane  
corresponds to a particular case of 
(\ref{bulk_metric}) with $\eta=\tilde\eta$
 and can 
be written as 
\beq
a(t,y)=a_0(t) \A(y), \quad n=\A(y), \quad 
\A(y)=   \cosh\mu y-\left(1+{\rho\over\sigma}\right) \sinh\mu|y|.
\eeq
The gravitational waves appear in a perturbed metric of the form 
\beq
ds^2=-n^2 dt^2
+a^2\left[ \d_{ij}+E_{ij}^{TT}\right]dx^i dx^j
+dy^2,
\eeq
where the  `TT' stands for transverse traceless.
The linearized Einstein equations yield a wave equation for  
 $E_{ij}^{TT}$ that is separable. 
Writing
$E_{ij}=\varphi_m(t) \E_m(y) \ e^{i \vec k. \vec x}\  e_{ij}$, where $m$ 
stands for  the mass (from the four-dimensional point of view), 
one finds that  the time-dependent part must 
satisfy
\beq
\ddot{\varphi}_m +3H_0\dot{\varphi}_m+\left[ m^2+{k^2\over
a_0^2}\right] \varphi_m =0\,, \label{varphieom}
\eeq
whereas  the $y$-dependent part obeys  
\beq
\H_m''+4{\A'\over\A}\H_m'+ {m^2\over \A^2}\H_m = 0\,.
\label{Heom}
\eeq
Like in the Minkowski case, the latter equation can be reformulated as 
a Schr\" odinger type equation, 
\beq
\label{SE}
 {d^2\Psi_m\over dz^2} - V(z)\Psi_m =-m^2 \Psi_m \,,
\end{equation}
 after introducing  the conformal 
coordinate $z=z_b+\int_0^y d{\tilde y}/\A({\tilde y})$ (with $z_b=
H_0^{-1}\sinh^{-1}(H_0/\mu)$)
 and defining $\Psi_m\equiv \A^{3/2}\H_m$.
The potential is given by (see  figure) 
\beq
V(z)= {15H_0^2 \over 4\sinh^2(H_0 z)} +
{\textstyle{9\over4}}H_0^2
- 3\mu\left[1+{\rho\over\sigma}\right] \delta(z-z_{\rm b}) \,.
\eeq
\begin{figure}
\psfig{figure=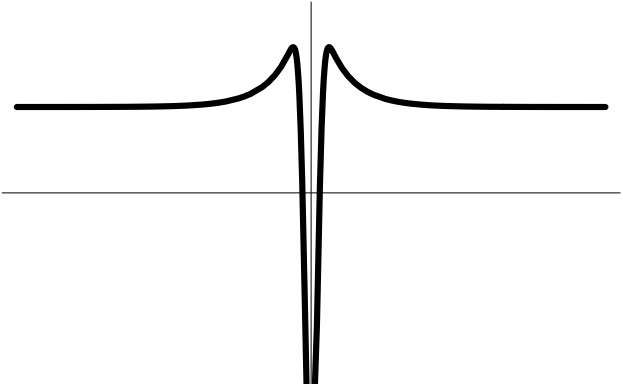,height=2in}
\caption{Potential for the gravitons 
\label{fig}}
\end{figure}
The non-zero value of the Hubble parameter implies the existence of  a gap, 
$\Delta m=(3/2)H_0$, between the zero mode and the continuum of massive 
Kaluza-Klein modes. The latter will decay during the inflationary phase, 
according to (\ref{varphieom}), 
leaving of relevance only the zero mode.  It  is simply  given by 
\beq
\H_0= C_1\equiv \sqrt{\mu}\,\,F\!\left({H_0/\mu}\right)\,,
\eeq
where, imposing the normalization $2 \int_{z_{\rm b}}^\infty |\Psi_0^2| dz=1$,
 the constant $C_1$ has been expressed in terms of $H_0$ via 
the function 
\begin{equation}
F\!\left(x\right) =\left\{ \sqrt{1+x^2} - x^2 \ln \left[ {1\over
x}+\sqrt{1+{1\over x^2}} \right] \right\}^{\!\!-1/2}
\!.\label{deff}
\end{equation}
Asymptotically, $F\simeq 1$ at low energies, i.e. $H_0\ll \mu$, 
and  $F\simeq \sqrt{3H_0/(2\mu)}$ at high energies, i.e. $H_0\gg \mu$.
One can then evaluate the vacuum quantum fluctuations of the zero mode 
by using the standard  canonical quantization. To do this 
explicitly, one writes the five-dimensional action for gravity at 
second order in the perturbations. Keeping only the zero mode and integrating
over the fifth dimension, one obtains  
\beq
S_{\rm g}= {1\over8\kappa^2}
\, \sum_{+,\times}\, \int d\eta\, d^3\vec{k}\,a_o^2\left[
\left({d\varphi_o \over
d\eta}\right)^2+k^2{\varphi_o}^2\right] \,, 
\eeq
This has the standard form for a massless graviton in
four-dimensional cosmology, apart from
the overall factor $1/8\kappa^2$ instead of $1/8\kappa_4^2$. It
follows that quantum fluctuations in each 
polarization, $\varphi_o$,
have an amplitude of $2\kappa (H_o/2\pi)$ on super-horizon scales. Quantum
fluctuations on the brane at $y=0$, where $E_o=C_1\varphi_0$, thus have the
typical amplitude
\beq
{1\over 2\kappa_4}\delta E_{\rm brane}=\left({H_0\over 2\pi}\right) F(H_0/\mu)
\eeq
At low energies, $F\simeq 1$ and 
one  recovers exactly the usual four-dimensional result 
but at higher energies the multiplicative factor $F$ provides an 
{\it enhancement of the gravitational wave spectrum amplitude 
with respect to the four-dimensional result}. 
An open question is how the gravitational waves will evolve during the 
subsequent cosmological phases, the radiation and matter eras. 

\section{Conclusions}
The idea of extra-dimensions hidden because ordinary matter/fields have not 
access to them has been very actively explored lately. We have presented 
here the simplest version of a cosmological scenario inspired by this idea.
At the level of homogeneous cosmology, the scenario seems so far consistent 
with observations since it deviates from standard cosmology only in the 
far past of the universe, for energy scales $\rho^{1/4}\gtrsim 10^3$  GeV, i.e. 
well before nucleosynthesis. Brane cosmological perturbations is a much more 
difficult question since one must take into account the propagation of 
gravitational waves in the bulk.

\end{document}